\documentstyle[12pt,aaspp4]{article}

\begin{document}

\title{Correlation Statistics of Irregular and Spiral Galaxies Mapped in \ion{H}{1} }

\author{E.E. Salpeter}
\affil{Center for Radiophysics and Space Research, Cornell University}

\and

\author{G. Lyle Hoffman}
\affil{Dept. of Physics, Lafayette College}

\begin{abstract}

Several measures of galaxy size and mass obtained from the neutral hydrogen mapping of 70 dwarf irregular galaxies (Sm, Im and BCD, hereafter generically called ``dwarfs'' irrespective of size or luminosity) presented in the preceding paper are compared statistically to
those for the set of all available \ion{H}{1}-mapped dwarfs and \ion{H}{1}-mapped spirals distributed within the same spatial volume to investigate variations in Tully-Fisher relations and in surface densities as functions of galaxy size and luminosity or mass.
Some ambiguities due to the ``non-commutativity'' of the correlations among the variables are addressed.
From linear regressions of logarithms we find that $\ell \propto r^{2.68} \propto v^{3.73} \propto m_H^{1.35} \propto m_{dyn}^{1.16}$ where $\ell$ is blue luminosity, $r$ is the geometric mean of the radius to the outermost detectable \ion{H}{1} and the optical radius, $v$ is the velocity profile half-width incorporating rotation and random motions, $m_H$ is the mass of \ion{H}{1}, and $m_{dyn} = v^2 r / G$. 
All are normalized by the values appropriate to a galaxy with blue luminosity of $10^9 {\rm L}_{\sun}$, typical of the region of overlap between dwarf and spiral galaxies.
The surface density $\Sigma_H$ of neutral hydrogen (averaged within the ``isophotal'' radius $r$) is almost constant along the sequence of size/mass/luminosity while surface density $\Sigma_L$ of blue luminosity increases with galaxy size.

For quantities not involving \ion{H}{1} we find no evidence for a ``break'' between dwarfs and spirals, but we do find some curvature in $\log v$ vs. $\log r$ and in the Tully-Fisher relation, $\log v$ vs. $\log \ell$.
Two consequences are:
{\it (i)} $\ell \propto v^{3.7}$ is more appropriate for the whole sequence than is $\ell \propto v^{2.5}$ as found for large spirals alone;
{\it (ii)} the surface density $\Sigma_{dyn}$ of total dynamic mass ($\propto v^2 / r$) is almost constant along the lower portion of the luminosity sequence, but increases appreciably with $\ell$ along the upper portion.

There is an indication for a difference in the correlations involving \ion{H}{1} mass or radius between dwarfs alone and spirals alone, in the sense that irregulars have somewhat more \ion{H}{1} mass or slightly larger \ion{H}{1} radii than spirals at a given blue luminosity, optical radius, or velocity profile width.
It is not clear if this is a true morphological effect or merely due to $m_H$ varying less strongly than $\ell$.
\end{abstract}

\keywords{Galaxies: Irregular --- Galaxies:  Kinematics and Dynamics --- Galaxies:  Structure --- Radio Lines:  Galaxies}

\section{Introduction}

In Paper I (Hoffman et al. \markcite{HSFRWH96} 1996) we have presented \ion{H}{1} mapping of a sample of 70 dwarf irregular galaxies distributed throughout the Virgo Cluster and the Local Supercluster.
Here and throughout this paper, we use ``dwarf'' to mean morphological type Sdm, Sm, Im or BCD irrespective of the intrinsic luminosity or size of the galaxy.
Other research groups have mapped dwarf galaxies in the meantime, and the available literature on mapped spiral galaxies has also been growing steadily.
In this paper we combine our sample of mapped dwarfs with those mapped in \ion{H}{1} by other authors using either single beam instruments or synthesis arrays, and with a sample of mapped spiral galaxies distributed within the same volume of space.
The combined sample is of interest for several questions:
Roberts \& Haynes \markcite{RH94} (1994) present a thorough study of how physical parameters vary along the morphological sequence.
Are dwarf irregulars, statistically speaking, just the low mass continuation of the spiral sequence, or a population with a distinct evolutionary history (Hodge \markcite{H89} 1989 and references therein; Gallagher \& Hunter \markcite{GH84} 1984; Fanelli, O'Connell \& Thuan \markcite{FOT88} 1988; Tyson \& Scalo \markcite{TS88} 1988; Dufour \& Hester \markcite{DH90} 1990; Schombert et al. \markcite{SBIM90} 1990; Westerlund \markcite{W90} 1990; Drinkwater \& Hardy \markcite{DH91} 1991; Kruger \& Fritze-von Alvensleben \markcite{KF94} 1994; van den Bergh \markcite{vdB94} 1994)?
Does the evidently episodic star formation history (Izotov, Thuan \& Lipovetsky \markcite{ITL94} 1994; McGaugh \markcite{M94} 1994; Meurer, Mackie \& Carignan \markcite{MMC94} 1994) of the dwarfs cause the Tully-Fisher correlations of optical properties (luminosity or diameter) vs. rotation velocity to deviate systematically from that for spirals (Hoffman, Helou \& Salpeter \markcite{HHS88} 1988; Schneider et al. \markcite{STMW90} 1990; Gavazzi \markcite{G93} 1993; Sprayberry et al. \markcite{SBIB95} 1995; Zwaan et al. \markcite{ZvdHdBM95} 1995)?
Do the unusual systems DDO 154, DDO 137 (Paper I) and \ion{H}{1} 1225+01 (Giovanelli, Williams \& Haynes \markcite{GWH91} 1991) continue to appear as unique systems when compared to a larger sample?

Recently a sequence of Low Surface Brightness (LSB) galaxies ranging in size from Malin I ,  comparable in size to the largest spirals but with much reduced star formation (past and current) across the disk, down to LSB irregulars of very small size, has been identified (Bothun et al. \markcite{BIMM87} 1987, \markcite{BSIS90} 1990; Impey \& Bothun \markcite{IB89}; Sprayberry et al. \markcite{SIIMB93} 1993; McGaugh, Schombert \& Bothun \markcite{MSB95} 1995; McGaugh \markcite{M95} 1995).
Another recent study (van Zee, Haynes \& Giovanelli \markcite{vZHG95} 1995) observed galaxies with particularly low blue luminosity to \ion{H}{1} mass ratios, which favors LSB galaxies.
Our Paper I selected dwarf galaxies by morphology, but does not necessarily favor LSB.
The combined sample of the present paper therefore covers a very large range in each of five extensive variables (blue luminosity $L_B$, radius $R$, velocity profile half-width $V_c$, hydrogen mass $M_H$, and indicative dynamic mass $M_{dyn}$) and of surface brightness $\Sigma_L = L_B / 4 \pi R^2$.
The selection criteria for our sample are given in detail in Sect. 2 (and selection biases are discussed in Sect. 3.1), but some salient features (besides the broad range covered) are:
\ion{H}{1} mapping was required for all dwarfs and spirals.
In addition to dwarfs (Sdm, Sm, Im and BCD), only late-type spirals (Sb through Sd) distributed within the same spatial volume were considered.
The Virgo cluster was included in the volume, but the omission of S0, Sa and Sab types and the insistence on a measured \ion{H}{1} radius (which selects against gas-stripped galaxies) means that the Virgo cluster core does not dominate our sample.

We selected a large and representative sample, but did not insist on high precision measurements.
Especially for the dwarfs, intrinsic variations as well as measuring errors are large, and some pairs of quantities are poorly correlated.
We therefore pay special attention to the ambiguities in the correlation statistics, following the discussion of Isobe et al. \markcite{IFAB90} (1990 --- hereafter IFAB).
Various measures of galaxy size and mass are available, and we select a particular set of five for analysis and discuss correlations among them in Sect. 4.
We address questions of ambiguity in Sect. 4.1 and questions of whether or not there are breaks between the dwarf and spiral sequences or curvature on log-log plots in Sect. 4.2.
We shall see that some of the ambiguity arises from ``non-commutativity'' in correlation statistics and from similar effects for products and ratios:
If some definition for regression lines gives $b \propto a^{\beta}$, $c \propto a^{\gamma}$, $(bc) \propto a^p$, then $p$ can sometimes differ appreciably from $(\beta \times \gamma )$.
Our data do not allow us to obtain local surface density profiles, but we have $\Sigma_L$, $\Sigma_H$ and $\Sigma_{dyn}$, the surface densities of blue luminosity, \ion{H}{1} mass and indicative gravitational mass averaged over the entire disks of the galaxies.
The correlations among these three ``intensive'' properties, and their variations with the extensive variables, are explored in Sect. 5.
The variation of the ratio $\Sigma_L / \Sigma_{dyn} = L_B / M_{dyn}$ is particularly interesting and particularly controversial.
We compare our analysis with that from four previous surveys (Gavazzi \markcite{G93} 1993; Roberts \& Haynes \markcite{RH94} 1994; Sprayberry et al. \markcite{SBIB95} 1995; Zwaan et al. \markcite{ZvdHdBM95} 1995) and discuss implications in Sect. 6.

\section{Sample Selection}

The definition of the sample we mapped at Arecibo is given in Paper I.
In brief, the galaxies are chosen from three lots:  45 from the {\it Virgo Cluster Catalog} (Binggeli, Sandage \& Tammann \markcite{BST85} 1985) and Leo (Ferguson \& Sandage \markcite{FS90} 1990); 14 from the field survey of Binggeli, Tarenghi \& Sandage \markcite{BTS90} (1990); and an additional 12 chosen to complete the sample of nearby dwarfs within 6 Mpc of the Sun mapped at Arecibo.
In combination, these subsamples produce a rather heterogeneous lot.

\markcite{BHK94}
\markcite{B82}
\markcite{BWKS80}
\markcite{Car85}
\markcite{CSS89}
\markcite{GS85}
\markcite{GWH91}
\markcite{Ha81}
\markcite{HGR79}
\markcite{HHS84}
\markcite{HST82}
\markcite{HHG83}
\markcite{HLHSW89}
\markcite{Hu79}
\markcite{HS85}
\markcite{HSM80}
\markcite{HSM81}
\markcite{HW79}
\markcite{KB84}
\markcite{Ro80}
\markcite{Sch89}
\markcite{SHST86}
\markcite{vZHG95}
\markcite{Beg89}
\markcite{BvdHA88}
\markcite{Bot89}
\markcite{BSvdK86}
\markcite{BV92}
\markcite{BCvD93}
\markcite{BK88}
\markcite{Bro92}
\markcite{BvW94}
\markcite{CB89}
\markcite{CBF90}
\markcite{CCBV90}
\markcite{CF88}
\markcite{CP90a}
\markcite{CP90b}
\markcite{CSvA88}
\markcite{CKBvG94}
\markcite{CvG92}
\markcite{CvG91}
\markcite{CvGBK90}
\markcite{CST94}
\markcite{CLV85}
\markcite{CCS91} 
\markcite{DvdH86}
\markcite{DHH90}
\markcite{Eng89}
\markcite{EGH90}
\markcite{GvGKB88}
\markcite{HDW86}
\markcite{HvWG94}
\markcite{IS91}
\markcite{ISTD87}
\markcite{IvD90}
\markcite{JC90}
\markcite{KB92}
\markcite{KDMB93}
\markcite{LSvG90}
\markcite{LS89}
\markcite{LV80}
\markcite{LS94}
\markcite{Lis92}
\markcite{LSY93}
\markcite{MCR94}
\markcite{OvdH89a} \markcite{OvdH89b}
\markcite{PHAU92}
\markcite{PFF+94}
\markcite{PC91}
\markcite{PCvG91}
\markcite{PCW91}
\markcite{PWBR92}
\markcite{Ran94}
\markcite{RDH94} 
\markcite{Rup91}
\markcite{SUP+90}
\markcite{SF93}
\markcite{SSL83}
\markcite{SC93}
\markcite{SS89}
\markcite{SB86}
\markcite{SBMW87}
\markcite{STTvW88}
\markcite{TY86}
\markcite{TBS93}
\markcite{TBPS94}
\markcite{TBFGSvW78}
\markcite{vABBS85}
\markcite{vDB91}
\markcite{vDvW94}
\markcite{vM88}
\markcite{Via90}
\markcite{VT83}
\markcite{War88a} \markcite{War88b} \markcite{War88c}
\markcite{WvdKA86}
\markcite{WAG93}

Additional dwarfs mapped in \ion{H}{1} were culled from the available literature (through the end of 1994).
All dwarfs mapped with sufficient resolution to determine \ion{H}{1} extents have been included, whether from single beam or synthesis array mapping.
We adopted for a sample of late-type spirals (types Sb through Sd) all those mapped similarly within 20 Mpc of the Local Group ($H_o = 75$ km/s/Mpc), excluding those thought to be members of tidally interacting pairs or groups.
Neither sample is complete in any sense (except that we have not rejected any non-interacting mapped galaxy {\it a priori}).
There is some bias toward systems in the Virgo cluster and its environs since so much Arecibo mapping effort has been expended in that area of the sky (Helou et al. \markcite{HGSK81} 1981; Helou, Hoffman \& Salpeter \markcite{HHS84} 1984; Hoffman et al. \markcite{HLHSW89} 1989), but the majority of both samples (dwarfs and spirals) is distributed throughout the Local Supercluster.
In all, including our 70 galaxies, we find a total of 112 mapped dwarfs out to a distance of 20 Mpc from the Local Group with another 16 at greater distances; the sample of mapped spirals numbers 119.
We drew the data from the original references (see Table 1) in each case. 

\placetable{tbl1}

In all cases, we have sought to obtain measures of galaxy properties equivalent to those for our Arecibo mappings as defined in Paper I:  blue luminosity $L_B$, optical radius at the 25 mag ${\rm arcsec}^{-2}$ isophote $R_{25}$, dynamical speed (rotation plus dispersion) $V_c$, galaxy-wide \ion{H}{1} mass $M_H$, \ion{H}{1} radius at $1/e$ of the central \ion{H}{1} intensity $R_{H,e}$ and \ion{H}{1} radius at the outermost detected point (at a sensitivity of typically a few $\times 10^{19}$ atoms ${\rm cm}^{-2}$) $R_{H,max}$.
Optical properties were taken from RC3 (de Vaucouleurs et al. \markcite{RC3} 1991) or equivalent photometry presented in the literature for each galaxy.
For the dynamically-relevant velocity we adopt $V_{c}^2 \equiv {V_{rot}}^2 + 3 {\sigma_{z}}^2$ where $V_{rot}$ is the inferred maximum rotation speed and $\sigma_{z}$ is the line-of-sight velocity dispersion averaged roughly over an Arecibo beam; see Paper I for discussion.
In the case of Arecibo or other single beam maps, we determined these quantities as in Paper I; for synthesis array maps we determined the quantities from the rotation curves provided by the authors, or from galaxy-wide (or single-beam) \ion{H}{1} profiles in a manner similar to that employed in Paper I.
Since inclination corrections to the rotation speed become quite uncertain at small inclination angles, and self-extinction corrections to the blue luminosity become uncertain for nearly face-on galaxies, we will mainly restrict the sample to $45\arcdeg \leq i \leq 85\arcdeg$ for types Sb through Sm.
Im and BCD galaxies of all inclinations are retained in the sample since it is not clear how inclinations should be measured for these galaxies in any case.

The diameter of the outermost recorded \ion{H}{1} intensity contour is taken for $D_{H,max}$.
For our Arecibo maps this point is assumed to be about half a beam-width in from the center of the outermost observed beam as explained in Paper I, and we defined $D_{H,max}$ similarly for other single-beam maps.
In the case of synthesis array maps for which a contour map was displayed, we simply measured the diameter of the lowest-level contour shown.
We defined $D_{H,e}$, the diameter at $1/e$ of the central flux, by fitting a flat-topped exponential model to the observed beam fluxes as detailed in Paper I for our Arecibo maps.
For synthesis array maps this quantity could be measured directly from the displayed contour maps by identifying the contour at a level $1/e$ times the central contour level.
Single-beam maps by other authors gave us more difficulty in defining $D_{H,e}$.
In some cases, e.g., van Zee, Haynes \& Giovanelli \markcite{vZHG95} (1995), a diameter defined in a different way turned out to be close to our $D_{H,e}$ for several galaxies in common between us and those authors, and so we simply adopted their definition as equivalent to ours.
For the several galaxies in the sample for which maps did not extend to the sensitivity limit we have determined $D_{H,max}$ from the regression of $D_{H,max}$ vs. $D_{H,e}$.
In this paper we shall use $D_{H,max}$ rather than $D_{H,e}$ since the former is more directly related to the total mass of the galaxy.
Unfortunately not all galaxies are measured to the same sensitivity limit; in particular, many of the early synthesis array maps do not reach below a sensitivity of $10^{20}$ atoms ${\rm cm}^{-2}$.
This introduces considerable systematic error into the measurement of $D_{H,max}$ in the sense that a number of the values are too low.

\section{Selection Biases, Extensive Variables and Comparison of Samples}

\subsection{Selection bias and hydrogen radius}

The selection criteria stated above introduce a selection bias in some (but not all) properties.
There is no bias against low optical surface brightness $\Sigma_L$ (if one discounts objects with such extremely small $\Sigma_L$ that special detection techniques are required), and our sample ranges over a factor of $\sim 10^3$ in $\Sigma_L$.
There is considerable bias against unusually small hydrogen mass $M_H$ for galaxies selected for hydrogen mapping.
There is no direct bias against galaxies with very small hydrogen radius $R_H$, i.e., a galaxy is included (as long as mapping was attempted) even if it has only an upper limit for $R_H$.
However, since hydrogen surface density $\Sigma_H \propto M_H / {R_H}^2$ tends to lie in a narrow range, the absence of very small $M_H$ indirectly discriminates against very small $R_H$.
The $M_H$ bias has two effects:
{\it (i)} galaxies of any luminosity $L$, which are greatly gas-deficient due to ram-pressure or tidal stripping, are likely to be missing, and 
{\it (ii)} the lower end of the normal distribution of $M_H / L$ is depleted for small $L$ but not for large $L$ (for indirect estimates of this effect see Sects. 4.2 and 5).

As discussed in the preceding section, we have attempted to obtain a value for the ``isophotal \ion{H}{1} radius,'' $R_{H,max} = \frac{1}{2} D_{H,max}$ for each galaxy in our own sample from Paper I and in the larger sample from the literature in Sect. 2 of this paper.
For a few galaxies in the combined sample we have only an upper limit to $R_{H,max}$ and in those cases we have taken the measurement to be half the formal upper limit.
In some cases this will be an overestimate of the true radius.
On the other hand, for a few galaxies in the sample drawn from the literature the adopted \ion{H}{1} radius may be too small as discussed in Sect. 2.
For all statistical purposes we shall omit the four ``special'' galaxies of Paper I along with HI 1225+01 (Giovanelli \& Haynes \markcite{GH89} 1989) which has a uniquely large \ion{H}{1} to optical radius ratio as we shall document below.
On the other hand, our omission of galaxies with particularly large $R_H$ is offset by the indirect bias against small $R_H$ mentioned above, i.e., the tails on both sides of the distribution in $R_H / R_{opt}$ are depressed somewhat.

The optical isophotal radius $R_{25}$ is known quite reliably for all the galaxies in our two samples, and Figure 1 displays $R_{H,max}$ vs. $R_{25}$ for each galaxy in the full sample.
(a) In the Figure, galaxies from Paper I are shown with larger symbols than those from the literature.
The inclusion of spirals in the sample flattens the slope of the best-fit correlation slightly (although it is still within the uncertainty of the slope for the sample of Paper I alone), but there is no evident systematic difference between the radii presented in Paper I and those obtained from the literature.
(b) As discussed in Paper I, many of the galaxies in our sample are drawn from the {\it Virgo Cluster Catalog} (Binggeli, Sandage \& Tammann \markcite{BST85} 1985) and some of those are within the core of the cluster, $< 5\arcdeg$ from its center.
Similarly, some of the mapped spirals drawn from the literature are in the Virgo cluster core.
While it is well-documented that some cluster spirals (Haynes, Giovanelli \& Chincarini \markcite{HGC84} 1984) and irregulars (Hoffman, Helou \& Salpeter \markcite{HHS88} 1988) are highly deficient in \ion{H}{1}, such galaxies are not favored by our selection criteria and any effect of stripping is lost in the overall scatter of the points in Fig. 1.
(c)  Fitting the 116 spirals separately from the 109 dwarfs, we obtain (using bisector regressions as discussed below) for spirals $R_{H,max} = (2.02 \pm .24) R_{25}^{(0.980 \pm .051)}$ and for dwarfs $R_{H,max} = (2.72 \pm .11) R_{25}^{(1.039 \pm .091)}$.
The powers are identical, within the uncertainties, but the fact that the scale factor is larger for dwarfs than for large spirals (by a factor of 1.4) is presumably due to two causes:
{\it (i)} $R_H / R_{opt}$ is overestimated, because of the selection bias, for small $L$ (dwarfs) but not for large $L$ (spirals), and
{\it (ii)} galaxies with an irregular morphology may have less efficient star formation and may have retained more of their gas (see also Sect. 4.2).
Linear regression on the logarithms of the radii (using the bisector of the direct and inverse regressions; see Sect. 4 and Isobe et al. \markcite{IFAB90} 1990) gives the correlation for spirals and dwarfs together:

\begin{equation}
\frac{R_{H,max}}{R_{25}} = (2.34 \pm .14) {\left( \frac{R_{25}}{4.25 \; {\rm kpc}} \right)}^{-0.110 \pm .031}
\end{equation}

\noindent
where the normalization length, 4.25 kpc, is a typical value for $R_{25}$ corresponding to $L_B = 10^9 L_{\sun}$.

\placefigure{radplot}

\subsection{Definitions of extensive variables}

To convert angular radii and magnitudes to physical units, we have taken distances given by a Virgocentric infall model with asymptotic $H_o = 74$ km $\rm s^{-1}$/Mpc and a Local
Group deviation from Hubble flow of 273 km $\rm s^{-1} \;$ toward Virgo, as in Paper I.
A number of the nearby dwarfs (especially those in the Local Group) have distance estimates from primary or secondary distance indicators; in those cases we have adopted the distances used in the sources of the \ion{H}{1} mapping with a few exceptions noted in Table 3 of Paper I.
The relative tightness of the correlation and its small deviation from linearity suggest that the hybrid surface densities (\ion{H}{1} mass over optical radius squared, e.g.) used by various authors differ by only a scaling factor from quantities derived from \ion{H}{1} measurements alone (with a few significant exceptions as indicated by the solid symbols in Fig. 1).
In later sections we prefer to use a single radius for correlations against other variables.
$R_{H,max}$ is in principle more appropriate since it gives the outermost point where we have velocity measurements, but $R_{25}$ is measured more reliably.
As a compromise, we shall use the geometric mean radius $R_{gm}$ between $R_{H,max}$ and $2.34 R_{25}$, so that $R_{gm}$ is close to $R_{H,max}$ in most cases:

\begin{equation}
R_{gm} = (2.34 R_{25})^{0.5}R_{H,max}^{0.5} .
\end{equation}

We shall find it convenient to normalize the various extensive quantities, namely luminosity, radius, velocity and \ion{H}{1} mass, by their values appropriate to a luminosity of $10^9 {\rm L}_{\sun}$.
Thus we define

\begin{equation}
\ell = \frac{L_B}{10^9 {\rm L}_{\sun}} , \;\;\;\;\;r = \frac{R_{gm}}{12.30 {\rm kpc}}, \;\;\;\;\;v = \frac{V_c}{80.51 \;{\rm km}\;{\rm s}^{-1}} , \;\;\;\;\;m_H = \frac{M_H}{5.76 \times 10^8 {\rm M}_{\sun}} .
\end{equation}

\noindent
For the indicative dynamical mass $M_{dyn} = {V_c}^2 R_{gm}/{\rm G}$ we then have 

\begin{equation}
m_{dyn} = v^2 r = M_{dyn}/1.212 \times 10^{10} {\rm M}_{\sun} .
\end{equation}

\noindent
Although $m_{dyn}$ is not an independent observable, but simply constructed from $v$ and $r$, we shall treat it separately so that we have five normalized extensive variables $\ell$, $r$, $v$, $m_H$ and $m_{dyn}$.
There are then 10 pairs of variables, and we display just 6 of the 10 in Figs. 2 and 3.

\placefigure{extbreak}

\placefigure{extcurve}

We will discuss the correlation statistics in detail in Sect. 4, but first we note how the galaxies from Paper I compare to the larger sample.
In each of the panels of Figs. 2 and 3, the Paper I data are shown with a larger symbol than data extracted from the literature.
The powers in the correlations above are all slightly smaller than those given in Paper I for our data alone, but always within the uncertainties ($3 \sigma$ rule) in the powers for the smaller dataset.
In most cases the flattening of the regression lines seems to be due mainly to the inclusion of the larger spirals in the sample, but in the case of $\log v$ vs. $\log \ell$ there is an apparent displacement of the points from Paper I with respect to those from the literature, in the sense that our sample has smaller rotation speed at a given luminosity.
This is most likely due to the selection by some authors (especially van Zee, Haynes \& Giovanelli \markcite{vZHG1995} 1995) of low surface brightness objects for mapping:
In terms of surface brightness $\Sigma_L$ defined below (Eqn. 9) their 11 galaxies (which survive our inclination cuts) have a mean of $\log \Sigma_L$ equal to ($-0.81 \pm .19$), compared with ($-0.22 \pm .05$) for the 58 surviving galaxies from Paper I.

\section{Correlation Statistics for Five Extensive Variables}

\subsection{The combined sample}

In the previous section we defined four physically independent variables, $\ell$, $r$, $v$, and $m_H$, which are ``extensive'' in the sense that they are related to overall size and mass.
One complication is that the measurement errors are not quite independent since $\ell$ and $m_H$ are proportional to $d^2$ and $r$ is proportional to $d$, where $d$ is the adopted distance which is sometimes controversial.
A second complication is related to $m_{dyn}$, defined in Eqn. (4), which is physically even more important as an extensive variable since total mass is less likely to fluctuate with time than $\ell$ does because of starbursts.
The mass $m_{dyn}$ can be treated as an independent variable for correlation purposes, but the results may be inconsistent with the fact that $m_{dyn}$ is also the product of $v^2$ and $r$.

Consider first the choices we would have if $m_{dyn}$ were truly independent.
Physically, we have to consider all five extensive variables on an equal footing {\it a priori}, so in each of the 10 pairs of variables we should choose one of the forms of correlation analysis that treats the variables symmetrically.
Variations in each variable are large, so that statistical results can depend strongly on whether the physical variables themselves or their logarithms are used.
Especially because of measuring errors, the logarithms of the variables are closer to being normally distributed than the variables themselves.
As in most previous analyses, we shall carry out all linear regressions using $\log_{10}$ of the variables $\ell$, $r$, $v$, $m_H$, and $m_{dyn}$, so that raising a physical variable to a power $p$ merely rescales the logarithm by a multiplying factor $p$ (see Fig. 4 and Sect. 4b, however).
Since we are interested in the fundamental relationship between the quantities rather than predictions for one quantity given a value for another, we seek a regression that treats errors in the two variables symmetrically.
Isobe et al. \markcite{IFAB90} (1990 --- hereafter IFAB) have discussed in detail three different symmetric definitions of linear regression, have shown that they usually give different values for the slope $\beta$ of a line $y = \alpha + \beta x$ and have demonstrated that the ``bisector method'' (the chosen regression being the bisector of the direct and inverse ordinary least-squares regressions) is usually preferred.
This (logarithmic) bisector regression for the other four extensive variables vs. $\ell$ gives

\begin{equation} \left. \begin{array}{ll}
r = (1.002 \pm .033) \ell^{0.382 \pm .013}, &
v = (1.001 \pm .027) \ell^{0.276 \pm .014}, \\
m_H = (1.028 \pm .075) \ell^{0.759 \pm .033}, &
m_{dyn} = (0.999 \pm .064) \ell^{0.859 \pm .027} .
\end{array} \right\}
\end{equation}

If the five variables were truly independent, a convenient way to proceed would be to keep the first power of one of the five physical variables and then to raise each of the other four to a particular power so that the rms deviation of the $\log_{10}$ from its mean, $\sigma_m$, is the same for all five.
We choose to keep the first power of $\ell$, for which $\sigma_m = 0.984$ in $\log_{10}$ with the mean corresponding to $\ell = 0.86 \times 10^9 {\rm L}_{\sun}$ (That $\sigma_m$ is close to unity is purely a coincidence).
The powers that give exactly $\sigma_m = 0.984$ for the other four variables would be 2.639 for $r$, 3.683 for $v$, 1.322 for $m_H$, and 1.165 for $m_{dyn}$, and for those choices we would have unit slope for all 10 linear regressions no matter which of the three symmetric definitions  we employ.
This would be convenient, but leads to an appreciable inconsistency in $m_{dyn} = v^2 r$, which is an example of the ``non-commutativity'' discussed in the Introduction:
If $r^{2.639}$ and $v^{3.683}$ were exactly equal to $\ell$, then $v^2 = \ell^{0.543}$, $r = \ell^{0.379}$, and $(v^2 r)^{1.165} = \ell^{1.074}$.
We thus have a 7.4\% discrepancy in slope with the direct correlation between $m_{dyn}^{1.165}$ and $\ell$.

Because of this discrepancy we cannot derive unique correlations but choose to adopt compromise powers which decrease the discrepancy slightly, although slopes are no longer unity.
We choose

\begin{equation}
\ell,\;\;\;\;\;r^{2.68},\;\;\;\;\;v^{3.73},\;\;\;\;\;m_H^{1.35},\;\;\;\;\;m_{dyn}^{1.16}
\end{equation}

\noindent
and use the bisector regression between all 10 pairs.
The slopes all lie between 0.97 and 1.03, and the formal statistical errors of the slopes (as defined in IFAB) lie between 3\% and 10\%.
Note that systematic errors in the slopes could well be much larger.
The discrepancy in slope for $v^2 \times r$ vs. $\ell$ where $v^2$ and $r$ are taken from the first three entries in Eqn. (5) against the direct $m_{dyn}$ vs. $\ell$ slope is 5.48\%.
The zero-points in the 10 regressions are within $\pm 0.01$ (in $\log_{10}$) when the normalizations in Eqn. (3) are used, but systematic errors are likely to be much larger (although hopefully much smaller than $\sigma_m \approx 0.98$).

Although the slopes of the 10 regressions are all close to unity, the tightness of the correlations varies appreciably.
Two measures of this tightness are displayed in Table 2:
The numbers in brackets are the correlation coefficients (the geometric mean of the slopes of the direct and inverse ordinary least-squares regressions) while the numbers without brackets give the rms deviation $\sigma_{reg}$ of either variable from the regression line.
The various $\sigma_{reg}$ are to be compared with $\sigma_m \approx 0.98$.
The tight correlation between $m_{dyn}$ and $v$ is mainly due to their interdependence ($m_{dyn} = v^2 r$).
On the other hand, the surprisingly tight correlation between $m_H$ and $r^{2.68}$ is only partly due to the fact that errors in distance $d$ almost cancel ($m_H \propto d^2$, compared with $d^{2.68}$ for $r^{2.68}$); the physical correlation must also be quite tight.
The poorest correlation is seen between $m_H^{1.35}$ and $v^{3.73}$, but here we cannot tell which of two possible causes for the poor correlation is dominant:  (1) Distance errors matter, since one variable is independent of $d$ whereas $m_H^{1.35} \propto d^{2.70}$ (and similarly, inclination errors affect $v$ but not $m_H$), and (2) the physical correlation between $m_H$ and $v$ might itself be poor.

\placetable{tbl2}

\subsection{Possible nonlinearities or breaks between dwarfs and spirals}

The scatter in most variables is larger for the dwarf galaxies than for the regular spirals, partly due to genuinely larger fluctuations and partly due to warping and other asymmetry in the disks giving larger errors in inclination and hence in $v$.
Linear regressions carried out separately for the dwarf irregulars (Hubble types Sdm and later) and for the regular late-type spirals (Sb to Sd) have larger statistical errors (especially for the dwarfs) than for the combined sample.
We have nevertheless carried out these separate regressions for the 10 pairs of ($\log_{10}$ of) the 5 variables in Eqn. (6), to look for any possible differences in slope and/or zero-point between dwarfs and spirals.
Some of the results are displayed in the dashed lines in Figs. 1 and 2.
Although there is some overlap in luminosity between types Sd and Sdm, results would have been very similar if we had separated $\ell < 1$ from $\ell > 1$ rather than by morphological type.

For the correlations involving \ion{H}{1}, there seems to be a parallel displacement between the dwarfs and spirals, with the dwarfs being more \ion{H}{1}-rich:
For the hydrogen radius vs. optical radius relation in Fig. 1, in the vicinity of $R_{25} \sim 3$ kpc, the dashed lines give $\log_{10} R_{H,max}$ larger for the dwarfs by $(0.158 \pm .074)$ than for the spirals.
Similarly in Fig. 2a where $1.35 \log_{10} m_H$ is plotted vs. $\log_{10} \ell$, the dwarfs lie higher than the spirals by a zero-point difference of $(0.71 \pm .19)$, corresponding to a factor 3.4 in \ion{H}{1} mass.
The difference in $\log_{10} R_{H,max}$ corresponds to a factor 2.1 in ${\left( R_{H,max} / R_{25} \right)}^2$.
The slopes of the pair of dashed lines are almost equal to each other ($1.30 \pm .09$ for spirals, $1.22 \pm .13$ for dwarfs) but, because of the displacement, are both larger than for the combined single line ($1 \pm .03$).
The difference by a factor of 3.4 between dwarfs and spirals (or between faint and bright galaxies) is statistically significant only in a formal sense, and it is smaller than the variance in either group.
Furthermore, this difference is reduced (although only slightly) if one corrects for the selection bias, discussed above, which has caused some galaxies below the regression line on the left hand side of Fig. 2a to be omitted.
A downward correction factor of the convenient form $(\ell / 100 )^c$ (with positive $c$) should probably be applied to $m_H$ from our figures, but two arguments show that $c$ is small:
{\it (i)} Stavely-Smith et al. \markcite{SDK92} (1992) have carried out a dwarf-spiral comparison for $m_H$ vs. $\ell$ (without mapping) which had no selection bias on $m_H$, and their power law for $m_H$ as a function of $\ell$ is very similar to that in our Fig. 2a with no $c$-correction;
and {\it (ii)} the selection bias affects only the low end of the distribution in $m_H / \ell$, but not the upper envelope of the scatter diagram in Fig. 2a.
The slope of this upper envelope is not very different from our solid line, and we feel that $c$ in the multiplicative correction factor $(\ell / 100 )^c$ for $m_H$ could not be much larger than about $+0.07$.
Our main physical conclusions are then that (a) there is no strong, sudden break between dwarfs and spirals, but (b) the ratio $m_H / \ell$ decreases with increasing luminosity $\ell$, perhaps as steeply as $\ell^{-0.27}$ but probably more like $\ell^{-0.20}$.

Another correlation of interest is the Tully-Fisher (TF) relation, $3.73 \log v$ vs. $\log \ell$, shown in Fig. 3a, where we again fit bisector regressions for spirals and dwarfs separately.
A break in the zero-point would be expected if there were a physical distinction between dwarfs and spirals, such as massive neutrinos contributing to the dark halos of spirals but not to dwarfs.
A mere change in slope would be expected if the TF relation is continuous but has curvature, as has been reported before.
For spirals we obtain a slope of $(1.31 \pm .14)$, about $3 \sigma$ steeper than that for dwarfs, $(0.88 \pm .10)$.
The zero-points are almost the same, $(-0.21 \pm .14)$ for spirals and $(-0.10 \pm .11)$ for dwarfs, suggesting continuity plus curvature (a continuous change in slope).
Although the errors in slope are large for our dwarfs and spirals separately, the significance of the change in slope is strengthened by comparing our combined sample (for which the error is much smaller, but dwarfs and irregular galaxies are still emphasized) with previous work:  
Writing $\ell \propto v^s$, our total sample gives a slope $s = (3.73 \pm .19)$, while Pierce \& Tully \markcite{PT88} (1988) find $s = (2.74 \pm .10)$ for regular (bright) spirals alone.
For practical application of the TF relation, some authors use the direct regression for regular spirals alone, which gives an even smaller value of $s \sim 2.5$.
However, for a physical understanding of the Hubble sequence, our use of symmetric regression and inclusion of dwarfs is more appropriate, and our averaged slope of $s = 3.73$ is closer to the original TF suggestion of $s = 4$.

The regression of $\log v$ vs. $\log r$ in Fig. 3b also shows some curvature, so that the luminosity-radius relation in Fig. 2c has almost constant slope throughout, $\ell \propto r^{2.6}$ or $r^{2.7}$ (the small displacement is probably not significant).
For the three log-log relations shown in Fig. 3, the curvature can be represented by a two-component fitting formula, but the slopes and ratios of coefficients are very uncertain.
The dashed curves in Fig. 3 show the following particularly simple choices:

\begin{equation}
v^5 \approx 0.4 \ell + 0.2 \ell^2 ,
\end{equation}

\begin{equation}
v^2 \approx 0.5 r + 0.4 r^{2.5} ,
\end{equation}

\begin{equation}
m_{dyn} \approx 0.4 \ell^{0.6} + 0.4 \ell^{1.2} .
\end{equation}

The use of logarithms in linear regressions is convenient and almost necessary if variables have a log-normal distribution.
However, the implied ``logarithmic averaging'' may not be the physically meaningful procedure.
In Newtonian mechanics the Virial Theorem requires that velocity-related averages over an orbit be carried out for $v^2$ rather than for any other power or for a logarithm.
We illustrate this with the logarithmic $v$ vs. $\ell$ relation in Fig. 3a for the dwarfs, where the dashed curve approximates $v^{(4.22 \pm .48)} = (0.76 \pm .21) \ell$ at the dwarf end.
In Fig. 4 we have replotted the same data for dwarfs on linear scales for $v^2$ and $\ell^{1/2}$ and show the bisector regression line $v^2 = (-0.039 \pm .048) + (1.02 \pm .12) \ell^{1/2}$.
In principle, averaging $v^2$ over a log-normal distribution centered on the zero-point of the $3.73 \log v$ - $\log \ell$ regression could give $\langle v^2 \rangle$ significantly larger than the square of $v$ corresponding to the zero-point itself, but in fact with the variance 0.652 in $3.73 \log v$ we get $\langle v^2 \rangle = 1.22$, almost the same as the value 0.98 from the regression of $v^2$ vs. $\ell^{1/2}$.

\placefigure{VT}

\section{Three Surface Densities}

Three surface densities are of physical interest, namely $\Sigma_L = L_B / 4 \pi R_{gm}^2$ (related to the average surface brightness in the blue), $\Sigma_H = M_H / 4 \pi R_{gm}^2$, and $\Sigma_{dyn} = M_{dyn} / 4 \pi R_{gm}^2$.
These are related to the extensive variables in Eqn. (6) by 

\begin{equation}
\frac{\Sigma_L}{1.23 \;{\rm L}_{\sun} {\rm pc}^{-2}} = \frac{\ell}{r^2} , \;\;\;\;\; \frac{\Sigma_H}{0.709 \; {\rm M}_{\sun} {\rm pc}^{-2}} = \frac{m_H}{r^2} , \;\;\;\;\; \frac{\Sigma_{dyn}}{14.9 \; {\rm M}_{\sun} {\rm pc}^{-2}} = \frac{v^2}{r} .
\end{equation}

\noindent
We should emphasize that the radius $r$, which we use in Eqn. (10) and in the rest of this paper, is mostly an ``isophotal'' or ``isosurface-density'' radius.
If we could use some ``scale-length'' radius $r_{sl}$ instead (unfortunately unavailable where the mapping is coarsely resolved), the variation in $\Sigma_{L}$ and $\Sigma_H$ would be greater, since low / high central surface brightness leads to a smaller / larger ratio of $r / r_{sl}$ (although only by a logarithmic factor for an exponential disk).
Central values of surface brightness and surface density would presumably also have larger variations.

If the correlation coefficients of the pairs of extensive variables were unity (i.e., straight lines on $\log$-$\log$ plots and $\sigma_{reg} = 0$), and if the $m_{dyn}$ inconsistency discussed in Sect. 4 were absent, Eqn. (6) would give uniquely the dependence of the surface densities on the extensive variables.
In reality, the correlations are not very tight, and there is some ambiguity and some curvature.
To illustrate the uncertainties, consider first only the first four entries in Eqn. (6) and pretend that they represent exact equalities.
This would give 

\begin{equation} \left. \begin{array}{l}
\Sigma_L \propto r^{0.68} \propto \ell^{0.25} \propto v^{0.95} \propto m_H^{0.34} , \\
\Sigma_H \propto r^{-0.015} \propto \ell^{-0.006} \propto v^{-0.021} \propto m_H^{-0.007} , \\
\Sigma_{dyn} \propto r^{0.44} \propto \ell^{0.16} \propto v^{0.61} \propto m_H^{0.22} .
\end{array} \right\}
\end{equation}

\noindent
Instead, we can obtain the ``direct logarithmic regression'' for each of the three surface densities separately against each of the five extensive variables in Eqn. (6).
The five alternative regression slopes (and their average) are shown in Table 3 for each of the three surface densities, together with the formal statistical error.
The alternative slopes can differ by more than the formal errors, which illustrates a systematic artifact for poorly correlated quantities:
A ratio like $\Sigma_L \propto \ell / r^2$ has a larger slope in its regression against $\ell$ and a smaller slope against $r^{2.68}$ than in Eqn. (11), whereas the regression against a ``neutral'' extensive variable (such as $v^{3.73}$ or $\ell r^2$) is probably more reliable.

\placetable{tbl3}

To summarize our data so far on the variation of the surface densities along the mass sequence:
The blue surface brightness definitely increases steadily (without curvature) along the sequence, approximately as $\Sigma_L \propto (\ell r^2 )^{\epsilon}$, with $\epsilon \sim$ (0.13 to 0.16)..
For our combined sampled, $\Sigma_H$ varies very little along the sequence, but we saw in Sect. 4.2 that a small correction in the form of a multiplying factor $\sim (\ell / 100 )^{0.07}$ should probably be applied to $m_H$ and hence to $\Sigma_H$) because of the bias against faint galaxies with small $m_H$.
That there is a correction to $\Sigma_H$, but that it is small, can also be seen indirectly from the scatter diagram in Fig. 5a of $\Sigma_H$ vs. $\Sigma_L$ for our sample:
The selection bias applies only to the lower portion of the diagram, not to the upper envelope which has only a small positive slope.
Some of the positive correlation must be an artifact due to measuring errors in $r$, which enters both $\Sigma_H$ and $\Sigma_L$ to the same power.
Thus, even the corrected $\Sigma_H$ should increase little along the sequence.
The variations in the scatter diagram of $\log \Sigma_L$ vs. $\log \Sigma_{dyn}$ (Fig. 5c) combine systematic variations along the sequence with individual variations ``perpendicular to the sequence.''
If one attempts to describe the ``parallel'' variation (along the sequence) as $\Sigma_L \propto (\Sigma_{dyn})^{p_{\parallel}}$, then Eqn. (10) and Table 3 give $p_{\parallel} \sim 1.6$ or 1.8.
However, the inclusion of curvature in Eqn. (8) gives

\begin{equation}
\Sigma_{dyn} \propto (1 + 0.8 r^{1.5})
\end{equation}

\noindent
for the variation along the sequence, i.e., $\Sigma_{dyn}$ hardly varies at the low end of the sequence (whereas $\Sigma_L$ does), and $p_{\parallel}$ is not a very meaningful parameter.

To investigate the variation of $\Sigma_L$ against $\Sigma_{dyn}$ ``perpendicular to the sequence,'' we eliminate the variations along the sequence by using slightly altered definitions:
We choose

\begin{equation}
\Sigma_L^{\prime} = \frac{\Sigma_L}{(\ell r^2 )^{0.16}} =  \frac {{\ell}^{0.84}}{r^{2.32}} , \;\;\;\;\;\;\;\;\Sigma_{dyn}^{\prime} = \frac{2 \Sigma_{dyn}}{1 + 0.8 r^{1.5}} = \frac{2 v^{2}}{r + 0.8 r^{2.5}}
\end{equation}

\noindent
so that $\Sigma_L^{\prime}$ and $\Sigma_{dyn}^{\prime}$ have little systematic variation along the sequence.
The scatter diagram is shown in Fig. 5d and the bisector regression line gives

\begin{equation}
\Sigma_L^{\prime} \propto ( \Sigma_{dyn}^{\prime} )^{p_{\perp}} , \;\;\;p_{\perp} = {1.118 \pm .085}
\end{equation}

\noindent
Note that a similar slope would be obtained if there were little true correlation between $\Sigma_L^{\prime}$ and $\Sigma_{dyn}^{\prime}$ but large measuring errors in $r$.
The true value of $p_{\perp}$ is therefore more uncertain than indicated by the formal error in Eqn. (14).

\placefigure{Surfdens}

\section{Conclusion}

\subsection{Summary.}

We have analyzed the statistical relations of five extensive (size-related) observables for a large set of galaxies, with a broad range of absolute luminosity and surface brightness.
Only galaxies with \ion{H}{1} mapping were included, early-type spirals (S0, Sa, Sab) and ellipticals were excluded along with spirals thought to be members of tidally interacting pairs or groups. 
Faint dwarf irregular galaxies are more prominent in our sample than they would be in a magnitude-limited catalog.
These five observables are $\ell$ (a normalized blue luminosity $L_B$), $r$ (a normalized ``geometric mean'' radius $R_{gm}$, formed from an isophotal or isosurface-density \ion{H}{1} radius close to that of the outermost detectable \ion{H}{1} and the optical radius $R_{25}$), $v$ (a normalized velocity profile width $V_c$ which incorporates both rotation and random motion), $m_H$ (the normalized total mass $M_H$ of neutral atomic hydrogen), and $m_{dyn}$ (the normalized indicative total dynamic mass out to radius $R_{gm}$, defined as $M_{dyn} = V_c^2 R_{gm} / G$).
These quantities are normalized by $10^9 {\rm L}_{\sun}$, 12.30 kpc, 80.51 km ${\rm s}^{-1}$, $5.76 \times 10^8 {\rm M}_{\sun}$, and $1.212 \times 10^{10} {\rm M}_{\sun}$, values typical of the region in which dwarf irregular and spiral galaxies overlap.

For our complete sample we analyzed all Hubble types Sb through Im, which gives a particularly large dynamic range of the five extensive observables, which peak around Sbc (Roberts \& Haynes \markcite{RH94} 1994).
Of the 225 galaxies in the complete sample, four were anomolously \ion{H}{1}-rich, and these were omitted in all statistical analyses.
Because of a selection bias for \ion{H}{1}-mapping, extremely \ion{H}{1}-poor galaxies are missing from the sample.
As most previous authors have done, we correlated logarithms of the variables rather than the variables themselves.  
Since the five variables are on an equal footing physically {\it a priori}, we followed Isobe et al. \markcite{IFAB90} (1990, IFAB) in using the ``bisector'' prescription for a symmetric form of a linear regression line for each of the 10 pairs of 5 variables.
Because of the large scatter in the data, correlations suffer some ambiguity from ``non-commutativity,'' especially for the product $m_{dyn} = v^2 r$.
Nevertheless, we find that the regression slopes for all 10 regressions between the $\log_{10}$ of $\ell$, $r^{2.68}$, $v^{3.73}$, $m_H^{1.35}$, and $m_{dyn}^{1.16}$ lie between 0.97 and 1.03, with statistical errors $\lesssim 10$\%.
Because of the non-commutativity for ratios, we again have some ambiguities as to the variation along the size/mass/luminosity sequence for the surface densities $\Sigma_L = \ell / r^2$ for blue luminosity, $\Sigma_H = m_H / r^2$ for neutral hydrogen mass, and $\Sigma_{dyn} = v^2 / r$ for indicative dynamic mass.
The extent of the ambiguity and one definition of an average slope for each pair are given in Table 3.
We have also attempted to characterize two kinds of variations of $\Sigma_L$ with $\Sigma_{dyn}$ by first writing $\Sigma_L \propto (\Sigma_{dyn})^{p_{\parallel}}$ for the average changes along the mass sequence.
These two surface densities were then corrected for the average systematic variation along the sequence to obtain new variables $\Sigma_L^{\prime}$ and $\Sigma_{dyn}^{\prime}$.
Then we fitted a regression line of the form $\Sigma_L^{\prime} \propto (\Sigma_{dyn}^{\prime})^{p_{\perp}}$ to the corrected variables to represent the variation ``perpendicular to the mean sequence.''
The regressions gave $p_{\parallel} \sim 1.6$ or 1.8 and $p_{\perp} \sim 1.15$, but these present values are unrealistic for reasons given below.

We have looked for curvature in the logarithmic regressions of the various extensive variables against each other, and we also investigated whether or not there are any ``breaks'' in the correlations for ``dwarfs,'' defined as Sdm through Im (including BCD), irrespective of size or luminosity, vs. those for  ``spirals,'' types Sb through Sd.
For the variables not involving \ion{H}{1}, we find no significant breaks between Hubble types, but the large range of extensive variables enabled us to find curvature in some of the regressions:
For the Tully-Fisher relation $\ell \propto v^s$, the bright spirals (which are preferred for practical application) give $s \approx 2.5$ or 2.7, our total sample gives $s \approx 3.7$, and the fainter galaxies give $s$ slightly larger than 4, the power originally suggested by Tully and Fisher \markcite{TF77} (1977).
There is little curvature in the relation $\ell \propto r^{2.7}$, so that the surface brightness in the blue, $\Sigma_L = \ell / r^2$, increases steadily along the mass sequence ($\Sigma_L$ in the red or infrared increases even more rapidly).
There is curvature in the $\log v$-$\log r$ relation, so that the surface density of dynamic mass is almost constant at the low end of the mass/radius sequence and increases appreciably along the upper end of the sequence.
This variation, roughly $\Sigma_{dyn} = v^2 / r \propto (1 + r^{0.8})$, coupled with the steady progression of $\Sigma_L$, shows that a relation of the form $\Sigma_L \propto (\Sigma_{dyn})^{p_\parallel}$ is not very meaningful.
The concept of the ``perpendicular parameter'' $p_{\perp}$ in $\Sigma_L^{\prime} \propto (\Sigma_{dyn}^{\prime})^{p_{\perp}}$ is meaningful, but out present value for it ($\sim 1.15$) may be quite uncertain because of a coincidence involving powers of the galactic radius $r$, which may have large measuring errors at the moment:
The definitions of $\Sigma_L^{\prime}$ and $\Sigma_{dyn}^{\prime}$ are such that particularly large errors, in the absence of strong real correlation, could mimic our results.

Only for correlations involving \ion{H}{1} content are there tentative indications of ``breaks'' between dwarfs and spirals in the sense that the $R_H$-$R_{opt}$ relation lies higher by a factor of $\sim 1.3$, and the $m_H$-$\ell$ relation by a factor of $\sim 3$, for dwarfs than for spirals.
These ``displacements'' near $\ell \sim 1$ are smaller than the variances and it is not clear if the effect is real.
The surface density of \ion{H}{1} mass, $\Sigma_H$, in our sample varies little along the luminosity/mass/radius sequence, and the individual variations from $\Sigma_H \sim 0.7 \;{\rm M}_{\sun}\;{\rm pc}^{-2}$ are fairly small.
Particularly small values of $m_H / \ell$ and of $\Sigma_H$ tend to be absent (especially for dwarfs) because of a selection bias, and we have estimated that $m_H / \ell$ and $\Sigma_H$ should be corrected downwards from our results by a multiplying factor of order $(L_B / 10^{11} {\rm L}_{\sun} )^{0.07}$.

\subsection{Comparison with previous analyses.}

Four recent papers have carried out related correlation studies with slightly different emphases:
{\it (i)} Roberts \& Haynes \markcite{RH94} (1994) mainly investigated variations with Hubble type rather than with luminosity.
{\it (ii)} Gavazzi \markcite{G93} (1993) concentrated on variation with luminosity, as we do, but he included S0, Sa and Sab galaxies and did not have much data on very faint dwarfs or low surface brightness (LSB) galaxies.
On the other hand, {\it (iii)} Sprayberry et al. \markcite{SBIB95} (1995) and {\it (iv)} Zwaan et al. \markcite{ZvdHdBM95} (1995) present extensive data on LSB galaxies and compare with galaxies (of similar dynamic mass $M_{dyn}$) with normal surface brightness $\Sigma_L$.
While there is agreement on most trends, there is considerable controversy regarding the trends of $\Sigma_L$ (in the blue) and $\Sigma_{dyn}$ along the size / mass / luminosity / Hubble sequences and also on their correlations relative to each other ``perpendicular to the sequence'' (i.e., variations for fixed $m_{dyn}$ or $\ell$).
We show first that the apparent discrepancies are partly due to uncertainties brought about by some near cancellations for variations along the sequence:
Assume that $\ell = v^{\alpha}$ and $r = v^{\beta}$ along the whole mean mass sequence.
With $\Sigma_{dyn} = v^2 / r$ and $\Sigma_L = \ell / r^2$, we have

\begin{equation}
\Sigma_{dyn} = v^{(2 - \beta )} ; \;\;\Sigma_L = v^{(\alpha - 2 \beta )} ; \;\;p_{\|} = \frac{\alpha - 2 \beta}{2 - \beta} .
\end{equation}

\noindent
Our derived approximate values are $\alpha \sim 3.8$ and $\beta \sim 1.4$, so that $\alpha - 2 \beta$ is appreciably smaller than $\alpha$ and $2 - \beta$ smaller than 2.
This leads to a large uncertainty in $p_{\|}$ even for constant $\alpha$ and $\beta$.
Furthermore, we saw that $\alpha$ and $\beta$ vary along the sequence (with $2 - \beta$ essentially vanishing on the low end), so that $p_{\|}$ is not well defined.

Consider next the ``variations perpendicular to the sequence,'' i.e., changes in $\Sigma_{dyn}$ at fixed $m_{dyn} = v^2 r$, so that $v^4 \propto r^{-2} \propto \Sigma_{dyn}$.
Assume that $\Sigma_L \propto (\Sigma_{dyn} )^{p_{\perp}}$ for variation at constant $m_{dyn}$ with some value for the constant $p_{\perp}$, so that

\begin{equation}
\ell \propto (\Sigma_{dyn} )^{p_{\perp} - 1} , \;\;\;\ell / v^{\alpha} = (\Sigma_{dyn} )^{(p_{\perp} - 1) - \alpha /4} .
\end{equation}

\noindent
For the special case of $p_{\perp} = 1$, the luminosity / mass ratio would then not depend on $\Sigma_{dyn}$ variations.
For the special case of $p_{\perp} = 1 + \alpha / 4$ on the other hand, variations in $\Sigma_{dyn}$ away from the mean value along the sequence would not affect the Tully-Fisher $v$-$\ell$ relation at all.

Both Zwaan et al. \markcite{ZvdHdBM95} (1995) and Sprayberry et al. \markcite{SBIB95} (1995) compared the Tully-Fisher (TF) relation between $v$ and $\ell$ for normal galaxies with that for LSB galaxies, which gives some information on Eqn. (15).
However, there is some spread in the observations and the two papers emphasize different aspects:
The former notes that the difference in the two TF relations is {\em small}, so that $p_{\perp}$ should be close to $(1 + \alpha / 4)$, which is about 1.9 (in the blue and using our bisector slopes).
On the other hand, the latter paper notes that there is {\em some} difference in the TF relations --- in the sense that ``Malin I-like'' objects with very large $r$ and small $v$, small $\Sigma_{dyn}$, have anomalously small $\ell / v^{\alpha}$.
These few galaxies at least indicate that $p_{\perp} < 1.9$ (If $p_{\perp}$ were as small as 1, then $\ell / m_{dyn}$ would depend little on $\Sigma_{dyn}$, but $\ell / v^{\alpha}$ would have a strong dependence.).
Qualitatively, our own result of $p_{\perp} \sim 1.12$ is intermediate between 1 and 1.9, as it should be, but the combined data of Zwaan et al. \markcite{ZvdHdBM95} (1995) and Sprayberry et al. \markcite{SBIB95} (1995) suggest a slightly larger value for $p_{\perp}$ than ours.
We saw that the large present-day errors in radius make the value of $p_{\perp}$ uncertain.

There is only an illusion of a discrepancy regarding the ratio $\ell / m_{dyn}$, which Gavazzi \markcite{G93} (1993) has decreasing along the mass sequence whereas our averaged linear regression has $\ell / m_{dyn} \propto m_{dyn}^{0.16}$.
Our Eqn. (9) indicates ``regression curvature'' in the sense that $\ell / m_{dyn}$ increases along the sequence for small $\ell$ and then decreases slightly for large $\ell$ [Fig. 2d in  Roberts \& Haynes \markcite{RH94} (1994)
also indicates a peak in $\ell / m_{dyn}$ at an intermediate Hubble type].

\subsection{Discussion.}

An important negative result of our compilation is the absence of any sudden break between dwarf irregulars and regular spirals for variables not involving \ion{H}{1}.
In an overall dynamic sense, small dwarfs are thus a continuation of the spiral Hubble sequence.
Only the fact that the $R_H$-$R_{opt}$ and $m_H$-$\ell$ relations seem to lie higher for dwarf irregulars than for spirals might indicate that gas depletion by star formation is reduced by morphological irregularities.
However, even this fact may only be a manifestation of the tendency (discussed below) that hydrogen content decreases more slowly with decreasing mass than luminosity does.

Of the three surface densities discussed above, only the blue surface brightness $\Sigma_L$ increases steadily and appreciably along the mass/radius/luminosity/Hubble sequence.
A fourth density $\Sigma_*$ for total stellar mass is not measured explicitly for most of our sample, but should be correlated best with $\Sigma_L$ in the visible to near infrared.
We do not have $\Sigma_L$ at those wavelengths either, but $L$, and hence $\Sigma_L$, is known to increase more rapidly along the sequence at longer wavelength than in the blue.
Our uncorrected data has neutral hydrogen mass surface density $\Sigma_H$ almost constant along the sequence; the correction for selection bias would presumably have $\Sigma_H$ increase along the sequence, but only very slowly, and $\Sigma_H / \Sigma_*$ definitely decreases along the sequence from dwarfs to giants.
As a consequence, the total disk mass density $(\Sigma_H + \Sigma_* )$ increases slowly (and is gas dominated) along the lowest portion of the sequence and increases rapidly along the middle and upper portions of the sequence.
The density $\Sigma_{dyn}$ of total dynamic mass (mostly dark matter) is almost constant along the lower sequence and increases along the upper portion, so that the ratio $(\Sigma_H + \Sigma_* ) / \Sigma_{dyn}$ probably increases quite slowly along the sequence.
The small variation of $\Sigma_H$, compared with the large variation of $\Sigma_*$, fits in with the suggestion by Kennicutt \markcite{K89} (1989) that star formation proceeds rapidly when the gas surface density exceeds a threshold value:
Rapid star formation on the upper sequence (where the initial gas density $\Sigma_H + \Sigma_*$ is large) depleted the gas rapidly, but the rapid depletion did not occur on the lower sequence so that the present-day $\Sigma_H$ is almost constant along the sequence.

The small variation of $\Sigma_H$ in our Figs. 5a and b stems in part from our radius definition and from a selection effect:
As mentioned in Sect. 4, the use of isophotal radii $r$ (instead of scale-length radii $r_{sl}$)  decreases variations  in both $\Sigma_L$ and $\Sigma_H$, compared with central values (de Blok, van der Hulst \& Zwaan \markcite{dBvdHZ95} 1995, private communication).
Galaxies with very small $\Sigma_H$ are likely to have had low total $m_H$ and therefore not to have been chosen as candidates for mapping in \ion{H}{1}, so the lowest portion of Figs 5a and 5b may be absent from our sample.
However, this is not the case for upward fluctuations, and it is significant that only a few out of the 225 galaxies have an unusually large \ion{H}{1} content for their position on the mass/luminosity sequence.
This also meshes with the notion that an initially high gas surface density usually produces rapid star formation and eventually depresses $\Sigma_H$ to ``typical values.''

We also have some qualitative information on the variation along the mass sequence of two different volume densities:
{\it (i)} The volume density $n_H$ of H (and other gas) in the midplane of the galactic disk depends on $\Sigma_H \times \Sigma_{dyn}$ and inversely on the velocity dispersion $\Delta V_{\perp}$ normal to the plane.
Although $V_c$ is smaller for dwarfs than for large spirals, $\Delta V_{\perp}$ varies little, as does $\Sigma_H$. Since $\Sigma_{dyn}$ decreases slowly with decreasing size/mass/luminosity, $n_H$ also decreases slowly.
{\it (ii)}  For studying tidal effects on a galaxy, another quantity with the dimensions of a volume density is important, namely $n_{tid} = M_{dyn} \left( 3 / 4 \pi R_{gm}^3 \right) \propto \Sigma_{dyn} / R_{gm}$.
Using Eqn. (6), we see that this quantity is actually slightly larger for dwarfs than for spirals, varying roughly as $n_{tid} \propto \ell^{-0.26}$.
Tidal effects should therefore be less pronounced on dwarfs than on spirals.
{\it (iii)} Although dwarfs are more ``robust'' from the point of view of {\it (ii)}, galactic winds, ``blow-outs,'' etc., depend on escape velocity [$\propto ( \Sigma_{dyn} R_{gm} )^{1/2}$], and dwarfs are less robust from this point of view.

Since the extensive variables peak around Sbc, it would be of interest to explore the differences in correlations of the extensive variables and surface densities for three morphological groupings:  Sdm through Im (including BCD), Sbc through Sd, and S0 through Sb.
There are at present too few detailed \ion{H}{1} mappings of early-type spirals to allow this comparison between the three groupings, but mapping is proceeding at a great rate.
It would also be useful to make more detailed comparisons between galaxies with and without \ion{H}{1} mapping to obtain $p_{\perp}$ more reliably.
For discussions of the empirical $\ell$-$v$ Tully-Fisher relation one wants to know if $p_{\perp}$ is always close to $(1 + \alpha / 4)$; if so, that would be a partial explanation for the narrow width of the relation.

\acknowledgments
We acknowledge fruitful discussions and correspondence with G. Bothun, J. Charlton, E. Corbelli, E. de Blok, J. Dickey, G. Gavazzi, G. Helou, S. McGaugh, M. Roberts, E. Skillman, D. Sprayberry, T. van der Hulst and M. Zwaan.  This work was supported in part by US National Science Foundation grants AST-9015181 and AST-9316213 at Lafayette College and by AST-9119475 at Cornell.

\clearpage

\figcaption{
\label{radplot}
Plot of radius $R_{H,max}$ to the outermost measured neutral hydrogen vs. optical radius $R_{25}$ at the 25 mag ${\rm arcsec}^{-2}$ isophote.
Irregular galaxies from Paper I are shown as large exes with BCDs as large open squares and upper limits shown at one-half the formal limit as open triangles.
Mapped irregulars from the literature are shown as small exes or squares, similarly.
Members of interacting binary systems are shown as asterisks and excluded from all correlations.
Mapped spiral galaxies from the literature are shown as small dots.
Open circles mark irregular galaxies with high luminosity ($> 10^{10}\;{\rm L}_{\sun}$).
Five special cases are indicated with filled symbols and are omitted from all correlations:
DDO 154 by a circle, DDO 137 by a triangle, the NGC 4532 / DDO 137 complex by a hexagon, VCC 2062 by a diamond, and HI 1225+01 by a small square.
The solid line is the bisector of the two ordinary least-squares regression lines ($y$ vs. $x$ and $x$ vs. $y$).
The two dashed lines are bisector regression lines for the spirals and dwarfs separately.}

\figcaption{
\label{extbreak}
Logarithmic scatter plots and regressions of pairs of extensive variables, each normalized to the value corresponding to a blue luminosity of $10^9 \;{\rm L}_{\sun}$ as described in the text.
The powers (multiplicative factors in the logarithms) are explained in Sect. 4a.
Symbols are chosen as for Fig. 1.
The solid line in each case is the bisector regression line, while dashed lines are for dwarfs alone or spirals alone.
(a)  Neutral hydrogen mass $m_H$ vs. blue luminosity $\ell$.
(b)  $m_H$ vs. indicative dynamical mass $m_{dyn}$.
(c)  Geometric mean radius $r$ vs. $\ell$.}

\figcaption{
\label{extcurve}
Logarithmic scatter plots and regressions of pairs of extensive variables, each normalized to the value corresponding to a blue luminosity of $10^9 \;{\rm L}_{\sun}$ as described in the text.
The powers (multiplicative factors in the logarithms) are explained in Sect. 4a.
Symbols are chosen as for Fig. 1.
The dashed curve in each panel is an a smooth curve which approximately matches the regressions for dwarfs alone at one end and for spirals alone at the other.
(a)  Velocity profile half-width $v$ vs. blue luminosity $\ell$.
(b)  $v$ vs. geometric mean radius $r$.
(c)  Indicative dynamical mass $m_{dyn} = v^2 r$ vs. $\ell$.}

\figcaption{
\label{VT}
Scatter plot and regression of normalized velocity profile half-width squared vs. square-root of normalized blue luminosity for dwarfs alone on linear, rather than logarithmic, axes.
The solid line is the bisector regression for the displayed points. }

\figcaption{
\label{Surfdens}
Logarithmic scatter plots of neutral hydrogen surface density, $\Sigma_H = m_H / r^2$, vs. (a) optical surface brightness, $\Sigma_L = \ell / r^2$ and (b) dynamical mass surface density, $\Sigma_{dyn} = v^2 / r$, and of (c) optical surface brightness, $\Sigma_L = \ell / r^2$ vs. dynamical mass surface density, $\Sigma_{dyn} = v^2 / r$ and (d) reduced optical surface brightness, $\Sigma_L^{\prime}$, vs. reduced dynamical mass surface density, $\Sigma_{dyn}^{\prime}$, as discussed in the text (Sect. 5).
Symbols are defined as in Fig. 1.}

\begin{table}
\dummytable\label{tbl1}
\dummytable\label{tbl2}
\dummytable\label{tbl3}
\end{table}

\end{document}